# Spatial beam self-cleaning in multimode fibres

K. Krupa[1,2*], A. Tonello[1], B. M. Shalaby[1,3], M. Fabert[1], A. Barthélémy[1], G. Millot[2], S. Wabnitz[4], and V. Couderc[1]

[1] Université de Limoges, XLIM, UMR CNRS 7252, 123 Avenue A. Thomas, 87060 Limoges, France
[2] Université de Bourgogne Franche-Comté, ICB, UMR CNRS 6303, 9 Av. A. Savary, 21078 Dijon, France
[3] Physics Department, Faculty of Science, Tanta University, Egypt
[4] Dipartimento di Ingegneria dell'Informazione, Università di Brescia, and INO-CNR, via Branze 38, 25123 Brescia, Italy
e-mail* katarzyna.krupa@u-bourgogne.fr

**Multimode optical fibres are enjoying a renewed attention, boosted by the urgent need to overcome the current capacity crunch of single-mode fibre systems and by recent advances in multimode complex nonlinear optics [1-13]. In this work, we demonstrate that standard multimode fibres can be used as ultrafast all-optical tool for transverse beam manipulation of high power laser pulses. Our experimental data show that the Kerr effect in a graded-index multimode fibre is the driving mechanism for overcoming speckle distortions, leading to a somewhat counter-intuitive effect resulting in a spatially clean output beam robust against fibre bending. Our observations demonstrate that nonlinear beam reshaping into the fundamental mode of a multimode fibre can be achieved even in the absence of a dissipative process such as stimulated scattering (Raman or Brillouin) [14,15].**

Beam propagation in multimode fibres (MMFs) is subject to a complex interplay of spatio-temporal processes. However, only few studies addressed nonlinear pulse propagation in MMFs, leaving this field largely untapped for the past thirty years. Very recently, there has been a resurgence of interest in MMFs for both fundamental and applied research. MMFs could provide a solution to meet increasing demands of new breakthrough technologies for light control and manipulation in communications, high-power fibre lasers and metrology [1,2,16]. In fundamental physics, MMFs may provide a natural tool for investigating spatio-temporal soliton dynamics [5,6], and for unveiling new, exciting nonlinear phenomena [3,4,9,13].

It is well known that light experiences an inherent randomization when propagating along MMFs, whereby input laser beams of high spatial quality fade into irregular granularities called speckles. Fibre stress or bending, as well as technological irregularities of the fibre, couple different guided modes and introduce supplementary randomization of the transmission features. For this reason, MMFs are not ideally suited for beam delivery and were replaced by single-mode fibres (SMFs) since the early days of optical communications. Recent works demonstrated that specific signal-processing algorithms could be used to predict or manage the beam shape at the output of a MMF by controlling its input field [17-19]. In particular, the application of multiple-input, multiple-output (MIMO) digital signal processing techniques enables the use of spatial-division multiplexing based on MMFs [2]. For high-power beam delivery applications, the spontaneous recovery of spatial beam quality in MMFs has so far been experimentally achieved exclusively through nonlinear dissipative processes such as stimulated Raman scattering (SRS) [14] or stimulated Brillouin scattering (SBS) [20]. However, these techniques do not lead to any self-cleaning of the input laser beam. It is now known that for powers above a critical level (few MWs) self-phase modulation (SPM) may overcome diffraction for any size of the beam, and this may in turn collapse into a filament. Reference [21] proposed to use this effect for combining multiple laser beams into a single one. Recent experiments reported supercontinuum generation (SCG) of femtosecond light pulses in the anomalous dispersion regime of graded-index (GRIN) MMFs: the close resemblance with self-focusing and multiple filamentation suggested the potential role of GRIN MMFs for achieving beam cleanup through control of the input launching conditions [3,4].

In this Letter, we experimentally demonstrate that a single strong laser beam can display enhanced brightness when propagating through MMFs. We show a new regime of remarkably stable self-induced bell-shaped beam propagation in MMFs based on the nonlinear nonreciprocity of mode coupling that occurs in the normal dispersion regime.

In our experiments, we launched intense laser pulses with a temporal duration of 900 ps and a wavelength of 1064 nm into a standard 12-m-long GRIN MMF. Using a wide input Gaussian beam of about 40 μm in diameter, we simultaneously excited about 200 guided modes on both polarization components. Figures 1a-d show the transverse spatial beam intensity at 1064 nm as a function of the output peak power ($P_{P-P}$); the corresponding beam transverse profiles are displayed in Figs.1a'-d'. At relatively low power ($P_{P-P}$ = 3.7 W), the smooth input Gaussian beam evolved into an irregular transverse profile with several randomly distributed speckles at the output (Figs.1a,a'). As we increased the input power, the beam profile coalesced from a highly multimode to a well-defined bell-shaped structure in the core centre, surrounded by a low power speckled pedestal, as shown in Figs.1d,d' (see also Supplementary Movie 1). We note that we did not observe any significant change in the input/output power transmission ratio between the low and the high intensity excitation regimes. Importantly, this rules out the possibility of the beam self-cleaning effect being due to a higher level of nonlinear loss for the high-order modes (HOMs) relative to the fundamental mode. By performing a simultaneous spatial and spectral analysis with a dispersive spectroscope, and Young's double-slit experiments, we also verified that the redistribution of power towards a central quasi-Gaussian profile results neither from spectral broadening and loss of temporal coherence, nor from loss of spatial coherence. As presented in Fig.2a,b, spatial beam self-cleaning occurs with the frequency spectrum essentially unchanged. Whereas, as shown in Fig.2c,d, we obtained interference fringes with a similar contrast of 70%, when the two interfering parts of either the initial speckled beams, or of the self-cleaned beams, were spaced by a distance nearly equal to the beam diameter taken at half of the maximum intensity (FWHMI) (see also Supplementary Material). The observed behaviour resembles the effect of multimode wave condensation, which was theoretically predicted in Ref.[22] in the continuous wave limit, whereby an initial random distribution of guided modes exhibits an irreversible evolution towards an equilibrium state. In such a state, the fundamental mode grows up to a macroscopic level of occupation while remaining immersed in a sea of small-scale fluctuations. In Fig.3a, we illustrate by solid red circles the dependence of the output near-field beam FWHMI diameters on the output power for a fixed input beam condition leading to a largely extended speckle pattern at low powers. A similar behaviour was observed in the far-field region. To study the dependence of beam self-cleaning on the input conditions, for each power level, we measured the output beam diameter for a large, statistically significant number (1000) of different input beams (leading to different energy distributions among the guided modes). With reference to Fig.3b, we can see that at high powers there is a significant reduction in the average beam diameter, as well as a dramatic reduction in the corresponding standard deviation (see also Supplementary Material).

We also carried out numerical studies (see Supplementary Material), which indicate that the bell-shaped self-cleaned beam is essentially composed of the fundamental mode, and not of a coherent summation of few low-order modes. Indeed, simulations show that at large input beam powers the relative contribution of the fundamental mode to the output beam power is significantly enhanced, whereas the power contribution from other modes progressively decreases. This prediction is well confirmed by our experimentally measured self-cleaned beam diameter of 7.8µm FWHMI (see Figure 3a), which closely matches the theoretical value of the beam diameter for the fundamental mode of the considered GRIN fibre (7.45µm FWHMI) [23]. Note that the coalescence into the fundamental mode appeared at a power level of $P_{P-P}$ = 0.6 kW (see Figure 3a), which is well below the threshold for SRS (in our case, the initial formation of the first Raman Stokes sideband took place at $P_{P-P}$ = 5.6 kW) and below the catastrophic self-focusing threshold (~ MWs) [21]. In the spectral domain, the beam self-cleaning process was accompanied by a weak symmetrical spectral broadening of the pedestal, localized at more than 15dBs below the laser beam (see Fig.SM8a). Additionally, we observed that beam self-cleaning was very robust upon further manual squeezing and bending of the fibre (Supplementary Movie 2). This behaviour drastically differs from the common situation found in the linear regime, where the output beam is characterized by speckles that result from the linear interference of many excited modes (Supplementary Movie 2). It is important to note that linear coupling among guided modes was neglected in Ref. [22], whereas it is taken into account in our experiments. This shows that the beam self-cleaning effect is observed in situations that are considerably more complex than the framework under which the theory of classical wave condensation was formulated.

The observed self-induced spatial cleaning effect significantly enhances the output beam spatial quality, as indicated by the simultaneous concentration of energy in the near field and in the far field. To emphasize this effect, we measured the fraction of power at 1064 nm carried by the central part of the beam (selected with a diaphragm opened at a diameter of 12 µm) for increasing input powers, and compared it to the total output power at 1064 nm. The blue open squares in Fig.3a show that the power in the centre of the beam increases nearly linearly with the fibre output power up to the power threshold $P_{Th}$ ~ 0.5 kW, where an abrupt increment in slope by a factor 5.3 confirms the enhancement of the beam brightness. The linear behaviour following $P_{Th}$ indicates the establishment of a new equilibrium in the energy distribution between HOMs and the fundamental mode. This suggests that the process has a non-catastrophic nature, similar to multimode wave condensation [22]. By contrast, a nonlinear beam narrowing behaviour with increasing propagation distance is typically expected for a catastrophic effect such as self-focusing.

To get a further insight into the nonlinear mechanism at the basis of beam cleaning, we studied the evolution of the spatial beam profile along the MMF by gradually backward cutting the fibre. The experimental results obtained at $P_{P-P}$ = 44 kW are shown in Fig.4. Despite the relatively high power used in these experiments, the input Gaussian beam still splits into several guided modes along the first few hundreds of millimetres, giving birth to spatial speckles (Fig.4a). The observed scrambling of the initial phase front proves that, just after coupling light into the fibre, transverse beam evolution is predominantly affected by mode beating owing to their different wavenumbers. Going further along the fibre, our cut-back analysis reveals that a strong modification of the modal intensity distribution occurs: the brightness of the output beam increases in spite of the remaining weak random speckle background.

Spatial beam Kerr self-cleaning may be explained in terms of the nonlinear nonreciprocity of the mode coupling process [24]. In fibres with parabolic index profile, the propagation constants of the modes take equally spaced values, so that coherent mode beating induces a periodic local intensity oscillation along the fibre, which the Kerr effect translates into a periodic longitudinal modulation of the refractive index [4,10,12]. With a simpler two-mode excitation, such type of self-induced or dynamic long-period Bragg grating was previously exploited in [25] for demonstrating light-controlled mode conversion in a GRIN MMF. More generally, four-wave mixing (FWM) interactions involving pairs of beating modes at the same wavelength introduce Bragg-type quasi-phase-matching conditions for a variety of mode coupling processes. As a result, energy exchanges occur between the forward propagating fundamental mode and the HOMs. In the Supplementary Material we derive a simplified two-mode *mean-field* model that makes it possible to analytically describe the dynamics of the interaction between the fundamental mode and the HOMs, in terms of an effective coupling term, driven by the collective presence of all fibre modes through FWM. The mean field theory (MFT) allows us to study the behaviour of large complex models in terms of a much simpler model. With this approach, the interaction of a large number of modes with the fundamental mode is approximated as a single averaged effect. A key outcome of the MFT explaining the observed Kerr self-cleaning effect is the nonlinear nonreciprocal behaviour of the equivalent nonlinear coupler model. Nonlinear nonreciprocity arises precisely because of the presence of SPM, which is highly different for the fundamental and for the HOMs owing to the difference between their overlap integrals. Specifically, nonlinear nonreciprocal coupling behaviour manifests as follows. For a given and sufficiently large input total power, if the mode power distribution is initially in favour of the HOMs (see Fig.SM4d), then one observes periodic power exchange between the fundamental and the HOMs. On the other hand, if the initial power distribution is in favour of the fundamental mode (see Fig.SM4c), the beam power remains in the fundamental mode. This nonreciprocity is nonlinear in that it only manifests above a threshold power, a fact which is in qualitative agreement with our experiments. One can further speculate that, when extended to describe the collective interactions in the original complex multimode system, the nonlinear nonreciprocity of the MFT model smoothens the dependence upon initial conditions and leads to irreversibility of the energy flow into the fundamental mode. In Bose-Einstein condensates (BEC) [26], the macroscopic quantum self-trapping of an initial population imbalance occurs because of interatomic interaction in the Bose gas. The self-cleaning process may thus be seen as an analogous phenomenon for the ensemble of photons in a multimode fibre.

Our experimental observations are qualitatively well reproduced by numerical simulations performed by solving a generalized version of the nonlinear Schrödinger equation (GNLSE3D), comprising two transverse spatial coordinates *x,y* (to account for the spatial beam distribution), the time domain *t* and the propagation coordinate *z* [6,23,27-29]. Numerical simulations of spatial and spectral beam evolution along the fibre are displayed in Fig.5. Figs.5a-d show that the initially highly multimodal distribution of the laser beam inside the MMF is gradually 'attracted' towards a bell-shaped transverse profile, in agreement with experiments. The power of the fundamental mode increases by a factor ~ 2.3 after 1 m of propagation distance (see also Supplementary Material). To show that beam self-cleanup only results from Kerr dynamics and SRS does not drive it, we limited the peak power below the Raman threshold ($P_{P-P}$ < 5.6 kW) for most of the experiments presented here. Therefore, we did not include SRS in our simulations. For ease of comparison, in Figs.5a'-d' we also reported numerical simulations with relatively low intensity, *i.e.* in the linear regime (see Supplementary Movie 3).

In conclusion, we experimentally demonstrated light self-cleaning, which enables robust effective propagation of spatially bell-shaped beams in spite of the high number of permitted guided modes. Such observations may pave the way for developing novel photonic devices for a wealth of applications, based on combining an effective spatial single mode environment with large fibre core diameters. From a fundamental perspective, beam self-cleaning is similar to the condensation of multimode waves [22, 30], which is formally equivalent to the Bose-Einstein condensation of gas particles. Complex multimode dynamics may shed "new light" on other exciting physical phenomena including spatiotemporal solitons [6-8], spatiotemporal rogue waves and turbulence [31], and extend the synchronization of coupled nonlinear oscillators, as described for example by the Kuramoto model [32], to spatially extended natural systems.

Moreover, the presented results may stimulate new theoretical analysis of the complex interactions at play in nonlinear multimode systems.

**Acknowledgements**
K.K, A.T., B.M.S., M.F., A.B. and V.C. acknowledge the financial support provided by Bpifrance OSEO (Industrial Strategic Innovation Programme) and Horiba Medical (Dat@diag N°I1112018W), by Région Limousin (C409-SPARC) and ANR Labex SIGMA-LIM. S.W. acknowledges support by the Italian Ministry of University and Research (MIUR) (grants 2012BFNWZ2 and 2015KEZNYM), the Russian Ministry of Science and Education grant 14.Y26.31.0018, and the European Union's Horizon 2020 research and innovation programme under the Marie Skłodowska-Curie grant agreement No 691051. G.M. acknowledges support by iXcore research foundation, Photcom Région Bourgogne and ANR Labex Action. The authors thank F. Wise, L. Wright, Z. Liu and A. Picozzi for valuable discussions.

**Author contributions**
K.K. and V.C. carried out the experiments. A.T. and S.W. performed the numerical simulations and theoretical analysis. All authors analysed and interpreted the obtained results, as well as equally participated in the discussions and in the writing of the manuscript.

**Materials & Correspondence**
Correspondence and requests for materials should be addressed to K.K.

**Competing financial interests**
The authors declare no competing financial interests.

LIST OF FIGURES:

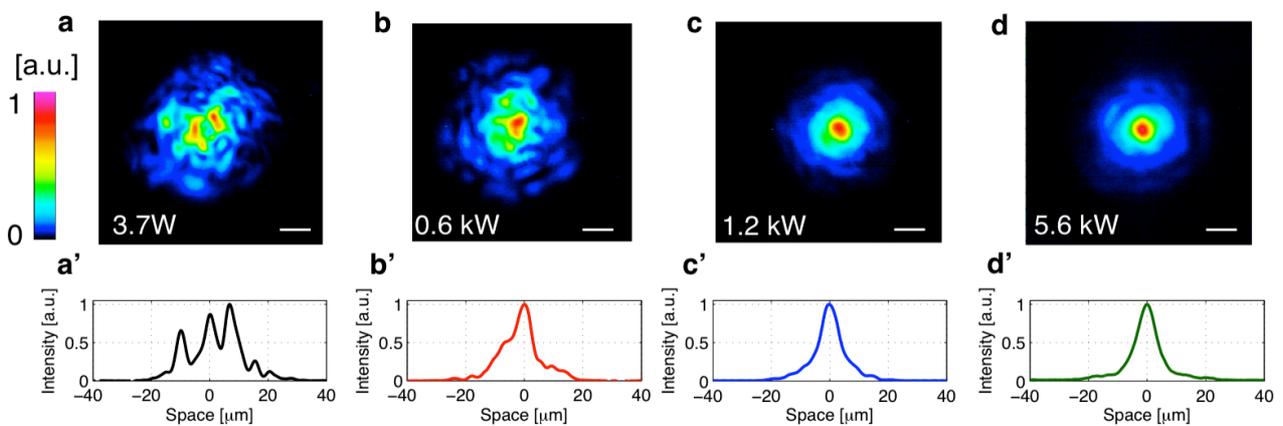

**Figure 1 Experimental nonlinear dynamics of beam self-cleaning in GRIN MMF. a-d**, Near-field images of the MMF output recorded at 1064 nm (intensities are referred to the local maximum and are shown in linear scale; scale bar: 10 μm) showing spatial beam self-cleaning, namely the formation of a well-defined bell-shaped transverse beam distribution when increasing the output peak power $P_{P-P}$. **a'-d'**,

Corresponding beam profiles versus x (y=0 section). Fibre length: 12 m.

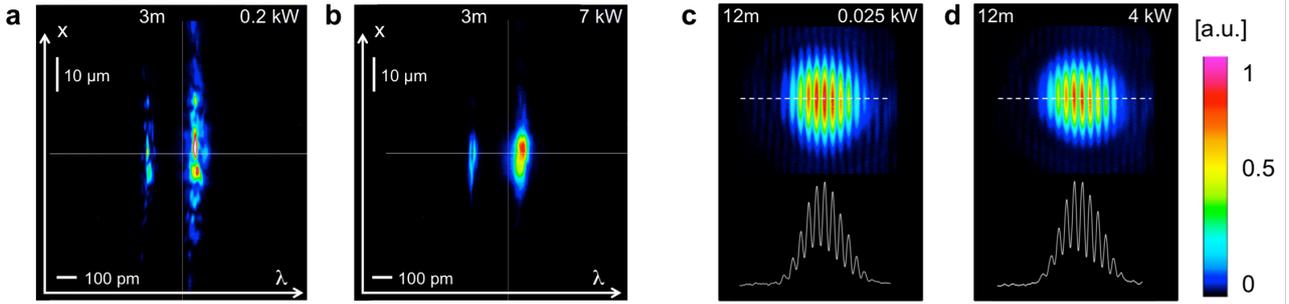

**Figure 2 Experimental analysis of temporal and spatial coherence of beam self-cleaning in GRIN MMF. a,b,** Spatio-spectral (x-λ) profile of output wave in the linear propagation regime at the output peak power $P_{P-P}$ = 0.2 kW (**a**), and in the beam self-cleaning regime at $P_{P-P}$ = 7kW (**b**). Fibre length: 3m. The output spectrum is essentially characterized by a nearly single longitudinal mode of the microchip laser with a weak second longitudinal mode. **c,d,** Examples of near-field interference patterns for the double-slit Young experiment; the corresponding x-axis cross section is obtained either in the linear propagation regime at the output peak power $P_{P-P}$ = 0.025 kW (**c**) or in the beam self-cleaning regime at $P_{P-P}$ = 4 kW (**d**). Fibre length: 12m.

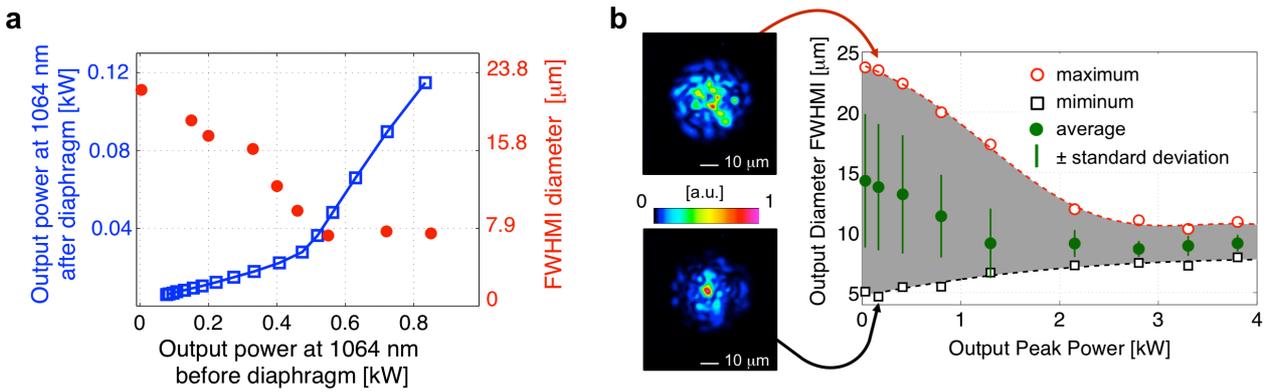

**Figure 3 Brightness enhancement and statistical analysis of beam self-cleaning in GRIN MMF. a,** Output power at 1064 nm measured in the central part of the fibre as a function of the output power (**blue open squares**), and corresponding FWHMI diameters of the near-field beam (at 1064 nm) (**red solid circles**). Brightness enhancement is demonstrated by the net increase of the blue curve slope. **b**, Average values of the FWHMI diameter (**green solid circles**) as a function of the output power, calculated from 1000 different input conditions, with the corresponding ± standard deviations (**green vertical lines**), and maximum (**red open circles**) and minimum (**black open squares**) diameter values. Red and black dotted curves are fits to guide the eyes. Insets: Examples of near-field images illustrating two extreme spatial beam distributions with maximum (top inset) and minimum (bottom inset) diameters recorded at low peak power of $P_{P-P}$ = 0.1 kW. Fibre length: 12m.

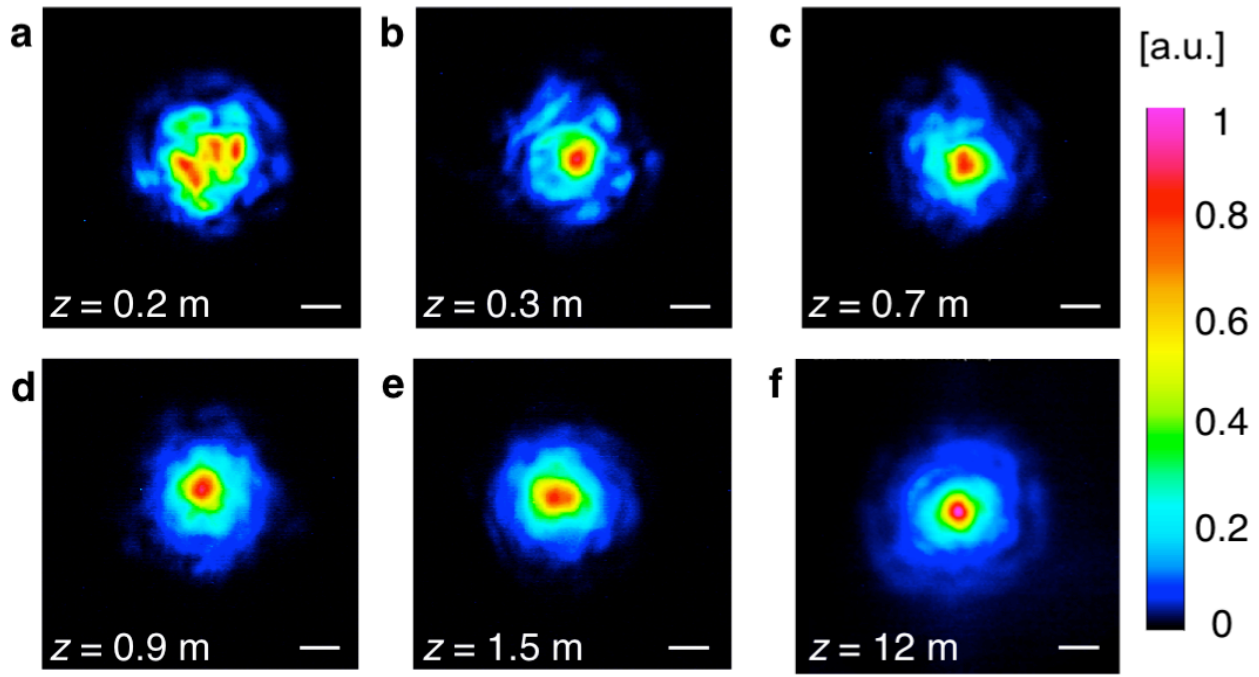

**Figure 4 Experimental cut-back results of beam propagation in GRIN MMF. a-f,** Near-field images recorded at 1064 nm (normalized intensity in linear scale), showing the development of beam self-cleaning along the propagation distance $z$. Output peak power at $z = 12$ m is $P_{P\text{-}P} = 44$ kW (*i.e.* high power). Scale bar: 10μm.

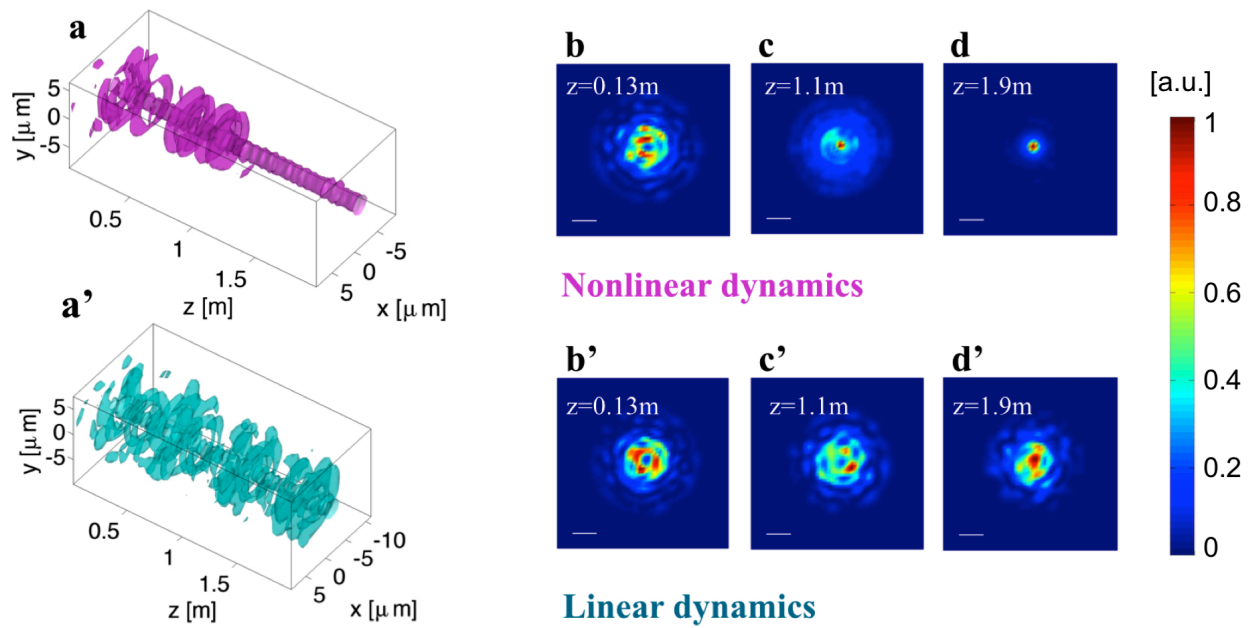

**Figure 5 Numerical results of beam propagation in GRIN MM fibre. a,a',** Spatial reshaping along the propagation distance $z$ showing the emergence of beam self-cleaning in nonlinear regime (**a**) whereas no self-cleaning occurs in the linear regime (**a'**). Iso-surfaces represent the points at 70% from the local maximum intensity value. **b-d,b'-d',** 2D output distributions in nonlinear (**b-d**) and linear (**b'-d'**) regime. Scale bar: 10 μm. Input peak power 63 kW (*i.e.* intensity $I = 5$ GW/cm$^2$). Spatial frames are obtained by averaging the time integrated intensity along $z$ over 3 consecutive samples (spaced by 5 mm).

## Methods

**Experimental setup.** We used a standard, commercially available GRIN MMF with 12 m length, 52.1 μm core diameter and 0.205 NA. The fibre was intentionally wound and bent in a fibre bundle in order to excite a large number of guided modes (~ 200), and also to investigate the influence of mechanical deformations (strong linear coupling) on beam self-cleaning. In GRIN MMFs, the maximum temporal broadening due to modal dispersion is expected to be in the range of 12 ps for 12 m of fibre. Thus in our experiments modal dispersion can be neglected. The laser source was an amplified Nd:YAG microchip laser, delivering 900 ps pulses at 1064 nm 30 kHz repetition rate. The polarized Gaussian laser pulses were launched into the fibre by using a lens with focal length of 50 mm and a three-axis translating stage. At the input face of the fibre the beam had a FWHMI diameter of 40 μm, which was close to the value of the fibre core diameter. The translating stage also allowed for controlling the initial spatial coupling condition, by adjusting the relative position between the fibre and the launched beam. We used an optical spectrum analyser covering the spectral range from 350 nm up to 1750 nm. The near-field beam profile (at the output face of the fibre) was imaged on a CCD camera through an 8 mm microlens with a magnification G=41. A 10 nm-wide bandpass interference optical filter at 1064 nm was introduced to analyse the spatial dynamics at the laser wavelength. Note that in our experiment, owing to the long pulse duration of the laser source and the low modal dispersion ensured by the parabolic profile of the fibre refractive index, a large number of initially excited modes could keep their temporal superposition, thus strong nonlinear coupling, for several tens of meters. To measure the fraction of power at 1064 nm carried by the central part of the beam represented by the blue open squares of Fig.2 we filtered the 82 times magnified output near field with a diaphragm of ~ 1 mm radius, which corresponds to a sensitive disc area of 12 μm diameter on the fibre exit face.

**Numerical Simulations.** We numerically solved the GNLSE3D, which is an envelope equation that stems from the Maxwell equations under the paraxial and unidirectional propagation approximations, with a nonlinear polarization limited to frequency components around the carrier wavelength of the laser beam. We used an integration step of 0.05 mm and a transverse 128x128 grid, for a spatial window of 150 μm x150 μm. We considered a standard GRIN MMF with core radius $\rho = 26$ μm and the maximum value for the core refractive index $n_{co} = 1.47$ and for the cladding index $n_{cl} = 1.457$. We took a truncated parabolic profile of the refractive index in the transverse domain $n(x,y)^2 = n_{co}^2(1-2\Delta r^2/\rho^2)$ for $r < \rho$ and $n(x,y)=n_{cl}$ otherwise, where $r^2=x^2+y^2$ and $\Delta=8.8 \times 10^{-3}$. For simplicity, we retained the frequency dependence of the silica refractive index only through the group velocity dispersion term at 1064 nm $\kappa'' = 16.55 \times 10^{-27}$ s$^2$/m, and we limited our numerical analysis to the scalar case, neglecting the polarization state of light. We considered a non-dispersive Kerr nonlinear index $n_2 = 3.2 \times 10^{-20}$ m$^2$/W and Raman fraction $f_R=0.18$ (in absence of Raman the Kerr nonlinear index was still reduced by a factor 1-$f_R$). We used an input beam diameter of 40 μm. To reduce computational time, we rescaled the problem by using an input intensity $I = 5$ GW/cm$^2$, (*i.e.* peak power 63 kW) which is higher than the one used in our experiments, so to reduce the propagation length to maximum $L = 2$ m, and we used the pulse duration of 7.5 ps. To mimic the speckled near-field image observed at low peak powers, we used two different approaches, which gave similar numerical results. The results presented in the main text of the Letter were obtained with the first technique, whereby we blurred an input flat spatial phase front with a random uniformly distributed 2D function and we solved the GNLSE3D equation with constant coefficients, as it would happen in an ideal fibre in absence of linear coupling among the guided modes. In this case the multiplicative input phase noise is intended to imitate the phase shifts in the first steps of propagation in the fibre. In the second approach, we applied a coarse-step distributed coupling method starting from a coherent input condition represented by a flat phase and Gaussian intensity distribution, and we perturbed the propagation by randomly changing selected key parameters, such as the fibre core diameter or the relative refractive index difference, and by introducing random artificial tilts to the beam in the fibre. We introduced these numerical perturbations every 1 mm, which is much longer than our basic integration step. Corresponding numerical results are reported in Supplementary Material.

**Data availability**
The data that support the plots within this paper and other findings of this study are available from the corresponding author upon reasonable request.